\documentclass[conference]{IEEEtran}
\IEEEoverridecommandlockouts
\usepackage{cite}
\usepackage{amsmath,amssymb,amsfonts}
\usepackage{algorithmic}
\usepackage{algorithm}
\usepackage{graphicx}
\usepackage{textcomp}
\usepackage{xcolor}
\usepackage{fancyhdr}
\usepackage{multirow}
\usepackage{multicol}
\def\BibTeX{{\rm B\kern-.05em{\sc i\kern-.025em b}\kern-.08em
    T\kern-.1667em\lower.7ex\hbox{E}\kern-.125emX}}
\begin{document}

\title{Knowledge Defined Networking for 6G: A Reinforcement Learning Example for Resource Management\\
}

\author{
    \IEEEauthorblockN{Erol Koçoğlu\IEEEauthorrefmark{1}, Mehmet Ozdem\IEEEauthorrefmark{2}, and Tuğçe BILEN\IEEEauthorrefmark{3}}
    
    \IEEEauthorblockA{\IEEEauthorrefmark{1}Department of Computer Engineering\\}
     \IEEEauthorblockA{\IEEEauthorrefmark{3}Turk Telekom, Istanbul,Turkey }
    \IEEEauthorblockA{\IEEEauthorrefmark{2}Department of Artificial Intelligence and Data Engineering\\Faculty of Computer and Informatics\\
    Istanbul Technical University, Istanbul, Turkey\\
    Email: kocoglu20@itu.edu.tr, mehmet.ozdem@turktelekom.com.tr, bilent@itu.edu.tr}
}

\maketitle

\begin{abstract}
6G networks are expected to revolutionize connectivity, offering significant improvements in speed, capacity, and smart automation. However, existing network designs will struggle to handle the demands of 6G, which include much faster speeds, a huge increase in connected devices, lower energy consumption, extremely quick response times, and better mobile broadband. To solve this problem, incorporating the artificial intelligence (AI) technologies has been proposed. This idea led to the concept of Knowledge-Defined Networking (KDN). KDN promises many improvements, such as resource management, routing, scheduling, clustering, and mobility prediction. The main goal of this study is to optimize resource management using Reinforcement Learning.
\end{abstract}

\begin{IEEEkeywords}
KDN, 6G, reinforcement learning, machine learning, resource management
\end{IEEEkeywords}

\thispagestyle{fancy}

\pagestyle{fancy}
\fancyhf{}
\fancyhead[C]{\scriptsize Accepted by International Conference on Computer Science and Engineering 2025 (UBMK), ©2025 IEEE}
\renewcommand{\headrulewidth}{0pt}

\section{Introduction}
The sixth generation (6G) of mobile communication networks is expected to redefine global connectivity by overcoming the limitations of 5G and supporting the rapidly growing demands of emerging digital ecosystems. As stated by Sing et al. \cite{6geval}, 6G targets unprecedented performance levels, including peak data rates over 1 terabit per second (Tbps), end-to-end latency below 0.1 milliseconds, and network reliability exceeding 99.99999\%. These advancements will enable next-generation applications such as real-time holographic telepresence, high-resolution extended reality (XR), autonomous systems, and underwater or aerial communications. Unlike earlier generations, 6G will feature deeply integrated artificial intelligence and machine learning across all architectural layers, enabling intelligent resource management, autonomous control, and predictive optimization. Architecturally, it envisions distributed and resilient sub-networks with hyper-specialized slicing and multi-path connectivity, potentially leading to a cell-less network paradigm. Sustainability is also a core principle in 6G, aiming to reduce the carbon and energy footprint of ICT infrastructure through energy-aware designs, efficient spectrum use, and real-time environmental sensing. Ultimately, 6G is expected to lay the foundation for the Internet of Everything (IoE), Smart Grid 2.0, Industry 5.0, and beyond—reshaping the interaction between society, digital technologies, and the physical world.

Small cells (SCs) are anticipated to play a key role in 6G, especially as the architecture moves toward ultra-dense, cell-free designs. As shown in the analytical study \cite{smallcell}, SCs help address major challenges such as load imbalance, coverage gaps, and uneven traffic distribution in IoT-dense scenarios. Their flexible, on-demand deployment within macro-cell areas improves network throughput, reduces latency, and enhances QoS fairness. Their proximity to users lowers path loss and interference, and increases energy efficiency by minimizing transmission power in both uplink and downlink. Moreover, SCs can adapt dynamically to shifting user densities making them a scalable and responsive solution to meet the high-capacity and dynamic requirements of 6G.

Despite its promise, realizing 6G requires overcoming several technical and systemic challenges. One major issue is the effective use of high-frequency bands like millimeter-wave and terahertz, which, while offering large bandwidth, suffer from limited range, atmospheric absorption, and poor penetration. Achieving ultra-low latency (microseconds) and extreme reliability (99.99999\%) across diverse applications such as autonomous systems and real-time XR demands a rethinking of current architectures and protocols. Energy efficiency is another concern, as infrastructure densification with massive MIMO and small cells may cause high power consumption if not optimized. Ensuring scalable and fair resource allocation in ultra-dense and heterogeneous environments—especially for IoT devices—is also critical. Furthermore, integrating AI and ML across the network stack introduces new complexities in algorithm design, data privacy, and decision accountability. According to Liao et al. \cite{keytech}, many current intelligent control methods lack effective evaluation, feedback, and optimization mechanisms. Most rely heavily on large-scale training data, making real-time performance and reliable convergence difficult. Addressing these multifaceted challenges is essential for the successful deployment and societal adoption of 6G technologies.

\begin{table*}[h]
\centering
\caption{Summary of Related Work}
\begin{tabular}{|p{2.6cm}|p{2.5cm}|p{2.8cm}|p{2.5cm}|p{2.8cm}|}
\hline
\textbf{Study} & \textbf{Focus} & \textbf{Technique} & \textbf{Strengths} & \textbf{Limitations} \\
\hline
DeepRM~\cite{b3} & Cluster resource scheduling & DRL (Deep Q-Network) & Efficient under heavy load & Limited scalability and interpretability \\
\hline
SO-KDN~\cite{b4} & Routing in SDN & RNN-based ML & High reliability, 90\% accuracy & Complexity, historical-data reliance \\
\hline
Xu et al.~\cite{b5} & Cloud-RAN energy management & DRL + Convex optimization & Reduced power, scalable decisions & Limited action flexibility, high DNN dependency \\
\hline
Zeman et al.~\cite{b6} & RL in KDN & Q-learning + P4 + INT & Real-time telemetry, adaptive control & General-purpose; lacks 6G-specific focus \\
\hline
ARTNet~\cite{b7} & SDV-F resource offloading & Multi-agent RL & Ultra-reliable, low-latency & Centralized load control limits resilience \\
\hline
Wang et al.~\cite{b8} & Vehicular DL allocation & DRL + closed-loop RRUI & Low-latency, reduced feedback & Sensitive to outdated local estimations \\
\hline
Sun et al.~\cite{b9} & Fog-RAN with AI capsules & DRL + centralized AI brain & Energy efficiency, distributed AI & Needs standardization and FL \\
\hline
Matera et al.~\cite{b10} & QoE/QoS in network slicing & ANN + regression models & QoE prediction, risk mapping & Simple models, scalability concern \\
\hline
\end{tabular}
\label{tab:relatedwork}
\end{table*}

Addressing these multifaceted challenges is essential for the successful deployment and societal acceptance of 6G technologies. In this paper, we propose a Knowledge-Defined Networking (KDN) architecture integrated with reinforcement learning (RL), particularly Q-learning, for dynamic resource management in 6G networks. Our main contributions are summarized as follows:

\begin{itemize}
    \item {Scalable Resource Management:} We develop a lightweight Q-learning agent capable of managing user association and base station (BS)/access point (AP) power control, addressing the scalability and heterogeneity issues in ultra-dense 6G topologies.
    \item {Energy Efficiency:} By dynamically adjusting transmission power and balancing user loads, the proposed model reduces overall network power consumption, aligning with 6G’s sustainability goals.
    \item {Latency-Aware Decision Making:} Our RL agent utilizes real-time telemetry (e.g., latency, packet loss) to make proactive decisions, helping to meet ultra-low latency and high-reliability requirements.
    \item {KDN with Real-Time Adaptability:} Through a unified KDN structure that includes SDN and telemetry-aware learning, our model continuously adapts to network dynamics, mitigating challenges in AI/ML integration such as feedback delay and real-time convergence.
    \item {Simulation Framework:} We present a simulation environment based on \texttt{ns-3} and OpenAI Gym via \texttt{ns3-gym}, enabling interaction between real-time network metrics and learning agents, and validating the effectiveness of our approach under realistic 6G configurations using the TeraSim extension.
\end{itemize}

The rest of the paper is organised as follows: Section II analyses the Related Works. Section III presents the proposed approach. Section IV evaluates the efficiency of the proposed model. Finally, the paper is concluded in Section V. 

\section{Related Work}
To address the unique demands of next-generation 6G networks, numerous research efforts have explored intelligent resource allocation, routing optimization, and adaptive control mechanisms using AI/ML techniques, particularly reinforcement learning (RL). These studies vary widely in scope, ranging from cloud resource management and SDN routing to vehicular networking and QoE prediction. Below, we highlight the most relevant contributions and position our approach in relation to them. Accordingly, Mao et al.~\cite{b3} present DeepRM, a deep reinforcement learning (DRL) framework for dynamic resource allocation in clusters. It replaces fixed rules with a neural network trained via reward signals that minimize delay and job completion time. DeepRM performs well under high loads by prioritizing smaller jobs, but suffers from issues related to training consistency, scalability, and interpretability. Unlike DeepRM’s simplified cluster setting, our work addresses 6G-specific challenges such as power control, mobility prediction, and dense device connectivity.\begin{figure*}[htbp]
\centerline{\includegraphics[width=0.55\textwidth]{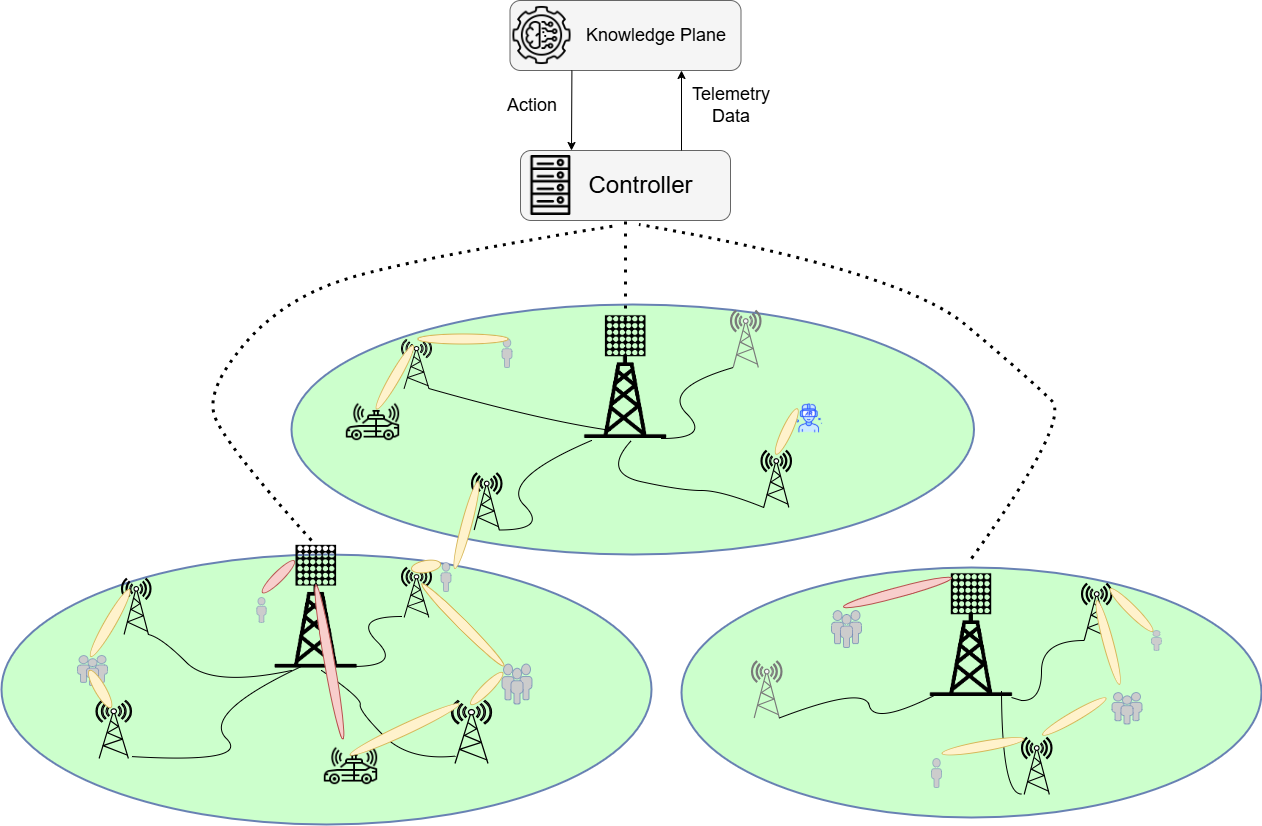}}
\caption{Our Topology Model}
\label{fig1}
\end{figure*} In~\cite{b4}, Ghosh et al. propose Self-Organised Knowledge Defined Networks (SO-KDN), an SDN-based architecture enhanced with an RNN for self-organized routing. The Most Reliable Route First (MRRF) algorithm selects paths based on historical reliability, reducing rerouting overhead. While promising in prediction accuracy and self-optimization, SO-KDN depends heavily on historical data and introduces architectural complexity. Unlike SO-KDN’s focus on routing reliability, our RL-based KDN approach targets dynamic resource management in 6G, especially in base station association and mobility control. Xu et al.~\cite{b5} apply DRL in Cloud-RANs for energy-efficient resource allocation. Their method activates/sleeps Remote Radio Heads (RRHs) via DRL and then uses convex optimization for active RRHs. It reduces power usage compared to traditional baselines. The model accounts for transition costs and combines ML with classical optimization. However, it allows only one RRH state change per step and depends on precise DNN training. Our approach considers multiple cells and is not limited to cloud-based RANs. Zeman et al.~\cite{b6} combine RL with KDN for autonomous network management. Using P4 and INT, the framework collects real-time telemetry for the Knowledge Plane to optimize network settings. Though both studies utilize RL and KDN, our work designs specialized RL algorithms for 6G resource optimization, offering improved suitability for ultra-dense networks. Ibrar et al.~\cite{b7} propose ARTNet, a multi-agent RL system for fog-based software-defined vehicular networks. It addresses high mobility and dynamic topology via distributed offloading and load balancing. Although ARTNet achieves low latency and energy use, its centralized load balancing may hinder resilience in highly distributed setups. In contrast, our study targets broader 6G scenarios beyond vehicular contexts. Wang et al.~\cite{b8} eliminate downlink feedback in vehicular networks using historical uplink RRUI data to drive DRL-based resource allocation. The model reduces latency but relies on timely and accurate RRUI, which may degrade performance under poor estimation. Their open-loop vehicular focus contrasts with our general-purpose 6G optimization framework. Sun et al.~\cite{b9} introduce an AI-driven Fog-RAN architecture with modular AI capsules for distributed intelligence, coordinated by a central AI brain. A DRL-based case study shows improved energy efficiency for computational offloading. They highlight challenges like interface standardization and federated learning. While they focus on fog RANs, our project implements RL over an ICCN architecture, but shares similar multi-scenario objectives. In~\cite{b10}, Matera et al. apply ML to predict QoE and QoS in network slicing. The mPlane-based architecture uses active and passive probes to monitor performance. Though ANN-based risk maps show superior prediction to regression models, the single-layer ANN may struggle with scalability. Our work instead uses RL as the core ML approach for adaptive decision-making.

As summarized in Table~\ref{tab:relatedwork}, prior work has made significant progress in applying RL and other AI techniques to various domains of network management. However, several gaps remain:

\begin{itemize}
    \item Many approaches focus on limited or specific domains (e.g., vehicular networks~\cite{b7,b8}, cloud RANs~\cite{b5}) and are not easily generalizable to the heterogeneous, ultra-dense environments of 6G.
    \item Some systems rely on historical data~\cite{b4} or simplified models~\cite{b10}, limiting their adaptability to real-time network dynamics.
    \item Several models do not integrate control decisions with simultaneous energy efficiency and latency considerations. Also, architectural complexity (e.g., reliance on custom APIs, offline training pipelines) may hinder scalability and practical deployment~\cite{b4}.

\end{itemize}

In contrast, our work offers a holistic and general-purpose solution tailored to 6G environments. By leveraging a KDN-based architecture and Q-learning, we support real-time telemetry-driven optimization that jointly considers throughput, energy consumption, latency, and user mobility across multiple cells. Our simulation results further validate the practicality of this approach under realistic 6G conditions.

\section{The Proposed Approach}
\subsection{KDN-Based 6G Topology}

The proposed model builds upon the Knowledge-Defined Networking (KDN) paradigm, which consists of network telemetry, SDN and ML \cite{array}. Network telemetry refers to the process of collecting, measuring, and analyzing data on a network's behavior and performance. It involves monitoring routers, switches, servers, and applications to understand their operations and track the flow of data across the network. The next component of KDN is SDN, which is a new approach in networking that separates the control plane (responsible for deciding how traffic is routed) from the data plane (responsible for forwarding traffic) \cite{sdn}. This decoupling enables centralized control and management, allowing network resources to be configured dynamically and flexibly to meet organizational requirements. The final part is Knowledge Plane (KP) with ML capabilities which can be said to be the brain of KDN.The idea of KP was first defined by Clark et al.\cite{b1} as a system within the network that builds and maintains high-level models of what the network is supposed to do. In the context of 6G, KDN is essential for managing increasingly dense, heterogeneous, and dynamic networks, such as those composed of both cell-free and cellular infrastructures. As illustrated in Fig. \ref{fig1}, in this paper, the KDN framework is implemented using a knowledge plane that collects telemetry data from the network environment and uses a learning-based agent to infer optimal control actions. The knowledge plane receives state information from the underlying network and responds with decisions regarding UE association and BS power control. This enables the system to operate autonomously, with continuous adaptation to environmental dynamics.

\subsection{Reinforcement Learning-Based Resource Management Model}

To address the dynamic and heterogeneous nature of 6G networks, we leverage reinforcement learning (RL) as the core intelligence for real-time resource management. Traditional machine learning models typically rely on offline training with historical datasets and require retraining to adapt to changes. In contrast, RL enables agents to learn optimal behaviors directly from interaction with the environment, offering a powerful mechanism for continuous adaptation under evolving network conditions. This is particularly crucial in 6G scenarios, where user mobility, traffic patterns, and interference levels vary rapidly and unpredictably.

\begin{figure}[htbp]
\centerline{\includegraphics[width=0.4\textwidth]{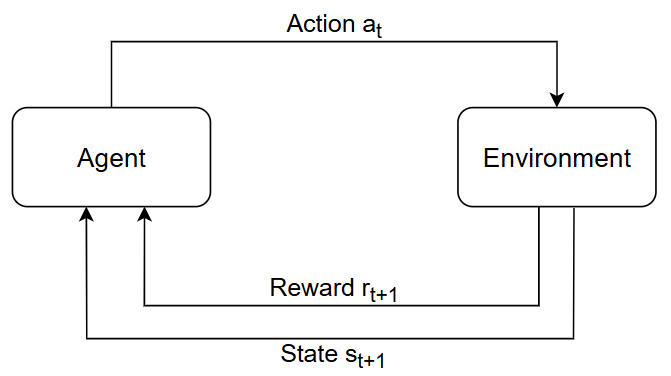}}
\caption{The interaction between the RL agent and its environment.}
\label{fig}
\end{figure}

As described by Raschka et al.~\cite{raschka}, the distinguishing feature of RL compared to supervised or unsupervised learning lies in its goal-directed exploration through a trial-and-error process. The agent observes the current state of the environment, takes an action, receives a reward signal, and updates its knowledge to improve future decisions. This closed feedback loop is shown in Fig.~\ref{fig}, and it forms the basis of online learning and control in our proposed KDN architecture.

Among various RL algorithms, we adopt \textit{Q-learning}, a model-free, off-policy method that is well-suited for discrete action spaces and low-complexity decision problems, such as user-to-cell assignment and transmission power control. Q-learning estimates the optimal action-value function \(Q(s,a)\), which represents the expected cumulative reward when taking action \(a\) in state \(s\), and subsequently following the optimal policy. The update rule for Q-learning is given in Eq. \ref{e1}.
\begin{equation}\label{e1}
Q(s,a) \leftarrow Q(s,a) + \alpha \left[ r + \gamma \max_{a'} Q(s',a') - Q(s,a) \right]
\end{equation}
Here, \( \alpha \) is the learning rate that controls the influence of new information. \( \gamma \) is the discount factor that balances immediate and future rewards. \( r \) is the reward obtained after taking action \(a\). Also, \( s' \) is the next state resulting from action \(a\).

As summarized in Algorithm \ref{a1}, the agent explores the environment using an $\epsilon$-greedy strategy, which balances exploration (trying new actions) and exploitation (using known good actions). This strategy is critical in dynamic environments like 6G, where the best action may shift as network conditions change. Also, the details of the components could be explained as follows:

\begin{algorithm}
\caption{Epsilon-Greedy Q-Learning Algorithm}
\begin{algorithmic}[1]\label{a1}
\STATE Initialize $Q(s,a)$ arbitrarily for all $s,a$
\STATE Set parameters $\alpha$, $\gamma$, and $\epsilon$
\FOR{each episode}
    \STATE Initialize state $s$
    \WHILE{$s$ is not terminal}
        \STATE Choose $a$ from $s$ using $\epsilon$-greedy policy
        \STATE Take action $a$, observe $r$ and $s'$
        \STATE Update $Q(s,a)$ using Equation (1)
        \STATE $s \leftarrow s'$
    \ENDWHILE
\ENDFOR
\end{algorithmic}
\end{algorithm}

\vspace{5pt}
\subsubsection{Agent}
The agent acts as the centralized intelligence within the Knowledge Plane of our KDN architecture. It continuously observes network telemetry and learns to perform control actions such as associating UEs to base stations or adjusting BS transmission power. Unlike traditional controllers with static rules, the agent evolves its policy based on feedback, making it resilient to changing traffic patterns and network topologies.

\subsubsection{State}
The state represents the current observable condition of the network. It provides the necessary context for decision-making and includes both user-specific and global features. In our implementation, the state vector includes:
\begin{itemize}
    \item {Packet Loss (\%)}: Indicates congestion or instability in the UE’s link.
    \item {Latency (ms)}: Average packet delay experienced by the UE.
    \item {Throughput (Gbps)}: Current data rate received by the UE.
    \item {UE Mobility Speed (m/s)}: Derived from location change rate.
    \item {UE Location}: Helps in selecting geographically optimal BS/AP.
    \item {Load Distribution}: Number of UEs connected to each BS/AP.
\end{itemize}
This multi-dimensional input allows the agent to capture both local (per-UE) and global (system-wide) context.

\subsubsection{Action}
Each action corresponds to a decision that directly affects network performance. In our model, possible actions include: (i) Assigning a UE to a different BS or AP, (ii) Adjusting the transmission power of a serving BS, (iii) Reallocating users across cells to balance load. The action space is discrete, making Q-learning particularly efficient in learning optimal mappings from state to action.

\subsubsection{Policy}
The policy determines the action selection strategy. During training, the agent uses an $\epsilon$-greedy policy to ensure exploration of suboptimal actions. As learning progresses, the agent increasingly exploits the learned Q-function, gradually converging to an optimal or near-optimal strategy.

\subsubsection{Value Function}
The value function, specifically the Q-function \(Q(s,a)\), guides the agent in evaluating the long-term benefit of actions taken in specific states. It acts as an internal scoring mechanism that allows the agent to anticipate the delayed effects of its choices—critical for long-horizon planning in dynamic systems.

\subsubsection{Reward}
The reward function quantitatively reflects the quality of an action. In our implementation, the reward is computed based on whether latency, packet loss, and throughput fall within predefined acceptable thresholds. Specifically, higher rewards are given when all metrics improve, and penalties are applied for SLA violations or inefficient load distributions. This formulation encourages the agent to favor balanced, energy-efficient, and delay-aware resource allocation strategies.

\section{Performance Evaluation and Results}
\begin{figure}[htbp]
\centerline{\includegraphics[width=0.45\textwidth]{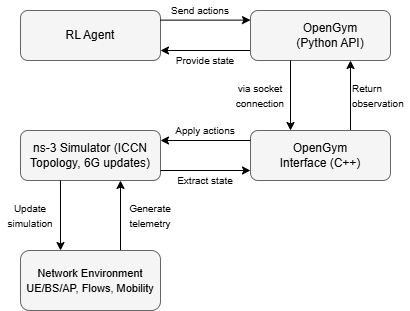}}
\caption{Our Simulation Model}
\label{fig2}
\end{figure}

For the performance evaluation, we utilize \textit{ns-3} (Network Simulator 3), a discrete-event network simulator widely used for academic and research purposes in the field of communication networks. ns-3 offers a flexible and modular architecture that allows researchers to simulate complex networking scenarios, including wireless communication, mobility models, and traffic generation.
 For this paper, we integrated ns-3 with the OpenAI Gym interface through the ns3-gym framework, allowing bidirectional communication between the simulation environment and our reinforcement learning agent. This setup enables the agent to observe network telemetry (e.g., packet loss, latency, and throughput) in real time and apply learned policies to optimize BS/AP selection in a simulated 6G network topology. For the physical capabilities of 6G, we used TeraSim extension \cite{thz} which enables THz level frequencies. The general structure of the simulation model is illustrated in Fig. \ref{fig2}.

In \cite{mallikarjun}, Mallikarjun et al. provides the performance metrics of the 5G SA network in different locations. The data shows that 5G can deliver high download speeds—up to 870 Mbps—and upload speeds exceeding 100 Mbps in optimal conditions, as seen at MP4. Latency remains relatively low, with the best performance reaching 7 ms at MP2, although some points exhibit slightly higher delays. While 5G demonstrates strong overall capability, there are noticeable variations in performance depending on location. This variation highlights the importance of network optimization and coverage. 

\begin{table}[h!]
\centering
\caption{5G Performance Results}
\begin{tabular}{|c|c|c|c|c|}
\hline
\multirow{2}{*}{\textbf{}}  & \textbf{Download (Mbps)} & \textbf{Upload (Mbps)} & \textbf{Latency (ms)} \\
\cline{2-4}
 &(Best / Avg.) & (Best / Avg.) & (Best / Avg.)  \\
\hline
\multirow{2}{*}{MP1}  & 707 & 83.8 & 8.5 \\
                      & 697.4 & 77.1 & 10 \\
\hline
\multirow{2}{*}{MP2}  & 283 & 70.2 & 7 \\
                      & 279.2 & 65.8 & 8.6 \\
\hline
\multirow{2}{*}{MP3}  & 474 & 51.6 & 9.5 \\
                      & 320.2 & 47.9 & 10.1 \\
\hline
\multirow{2}{*}{MP4}  & 870 & 110 & 8 \\
                      & 850.6 & 102.8 & 9 \\
\hline
\multirow{2}{*}{MP5}  & 456 & 36.4 & 9 \\
                      & 413.4 & 30.4 & 9.8 \\
\hline
\end{tabular}
\end{table}

\subsection{Our Simulation Results}
\subsubsection{Throughput}
 The following graph illustrates the aggregate network throughput (in Gbps) as the number of users increases, comparing the performance of the proposed RL-KDN system with a baseline Idle (non-RL) approach. As shown, both methods exhibit a general upward trend in throughput with increasing user count, indicating that the network scales with load. However, the RL-KDN model consistently outperforms the baseline across all user levels.This improvement is a direct result of the agent's ability to dynamically adjust UE-to-BS/AP associations and base station transmission power levels in real-time. At higher loads (e.g., 300 users), RL-KDN still maintains superior throughput, suggesting that the reinforcement learning agent effectively mitigates congestion and optimizes spectral efficiency. This demonstrates the system’s capacity to adapt to user density and deliver improved network performance compared to static, non-learning methods.
\begin{figure}[H]
    \centering
    \includegraphics[width=0.9\linewidth]{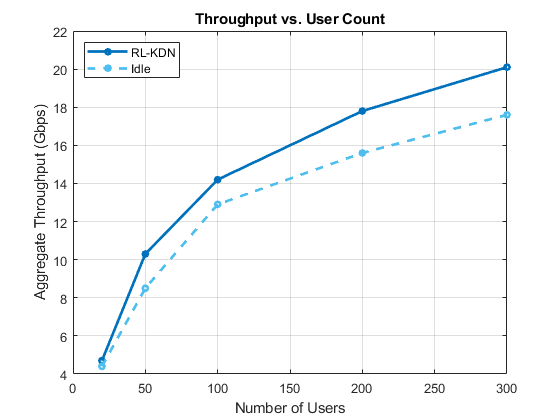}
    \caption{Throughput Results}
    \label{fig:enter-label}
\end{figure}

 \subsubsection{Latency}
 The graph below illustrates how end-to-end latency (in milliseconds) varies with increasing user load in both the proposed system and a baseline idle model. As user count grows from 20 to 300, both systems exhibit a rising latency trend due to higher contention and traffic congestion. However, our model consistently maintains lower latency.By dynamically reassigning users to less congested BSs or APs in response to changing traffic, our model effectively minimizes queueing delays and link overload.
  \begin{figure}[H]
    \centering
    \includegraphics[width=0.9\linewidth]{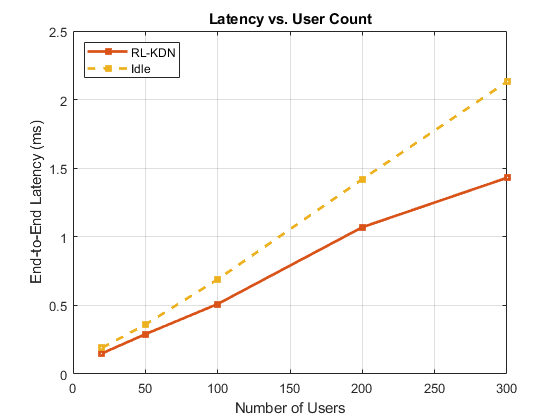}
    \caption{Latency Results}
    \label{fig:enter-label}
\end{figure}

\subsubsection{Packet Loss}
Our Q-learning model significantly reduces packet loss by continuously adapting to real-time network conditions. While many existing approaches react slowly to congestion or rely on fixed thresholds, our model proactively learns to distribute users across available base stations and access points based on observed packet delivery performance. 
\begin{figure}[H]
    \centering
    \includegraphics[width=0.9\linewidth]{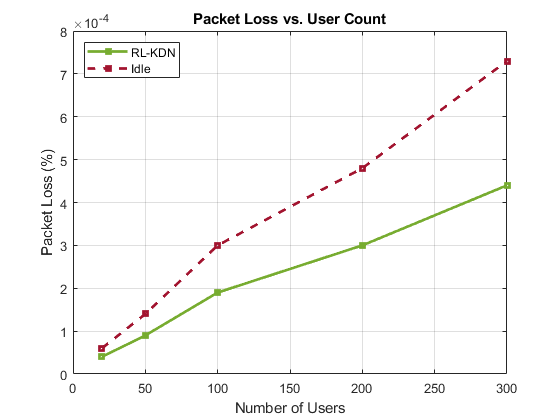}
    \caption{Packet Loss Results}
    \label{fig:enter-label}
\end{figure}

\section{Conclusion}
In this paper, we have explored the integration of KDN with RL, specifically Q-learning, as a promising approach to address the complex and dynamic resource management challenges anticipated in 6G networks. Performance evaluations demonstrate that our model consistently outperforms baseline systems in terms of throughput, latency, and reliability, particularly under high user densities. The proposed example highlights how real-time learning from network states can lead to efficient allocation strategies, improved performance metrics, and enhanced adaptability under uncertain conditions.

\end{document}